%% file: main.tex
\documentclass{article}

\usepackage{arxiv}

\usepackage[utf8]{inputenc} % allow utf-8 input
\usepackage[T1]{fontenc}    % use 8-bit T1 fonts
\usepackage{hyperref}       % hyperlinks
\usepackage{url}            % simple URL typesetting
\usepackage{booktabs}       % professional-quality tables
\usepackage{amsfonts}       % blackboard math symbols
\usepackage{nicefrac}       % compact symbols for 1/2, etc.
\usepackage{microtype}      % microtypography
\usepackage{lipsum}		% Can be removed after putting your text content
\usepackage{graphicx}
\usepackage[square, numbers, sort&compress]{natbib}
\usepackage{doi}
\usepackage{amsmath}
\usepackage{amsthm}
\usepackage{listings}
\usepackage{xcolor}
\usepackage[ruled,vlined]{algorithm2e}
\usepackage{multirow}
\usepackage[table]{xcolor}

\theoremstyle{definition}
\newtheorem{definition}{Definition}[section]

\theoremstyle{remark}
\newtheorem{notation}{Notation}[section]

\theoremstyle{remark}
\newtheorem{remark*}{Remark}[section]

\theoremstyle{definition}
\newtheorem{example}{Example}[section]

\lstdefinestyle{py}{
    language=Python,
    basicstyle=\ttfamily\small,
    keywordstyle=\color{blue!70!black}\bfseries,
    commentstyle=\color{gray}\itshape,
    stringstyle=\color{purple!70!black},
    showstringspaces=false,
    tabsize=4,
    breaklines=true,
    frame=single,
    captionpos=b
}

\SetKw{Or}{or}
\SetKw{And}{and}

\title{Meta Flip Graph meets Serendipitous Product: new Fast Matrix Multiplication results}

\author{\href{https://orcid.org/0000-0001-8047-0114}{\includegraphics[scale=0.06]{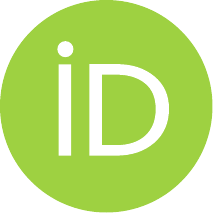}\hspace{1mm}Andrew I.~Perminov}\\
	Research Center for TAI\\
	Institute for System Programming\\
	Moscow \\
	\texttt{perminov@ispras.ru}
}

\renewcommand{\headeright}{}

\newcommand{\improvedCount}{207}
\newcommand{\rediscoveredZT}{84}
\newcommand{\betterStrassen}{23}
\newcommand{\betterStrassenTotal}{52}

\newcommand{\countZT}{374}
\newcommand{\countZ}{18}
\newcommand{\countQ}{288}

\newcommand{\percentZT}{55}
\newcommand{\percentZ}{2.65}
\newcommand{\percentQ}{42.35}

\hypersetup{
pdftitle={Meta Flip Graph meets Serendipitous Product: new Fast Matrix Multiplication results},
pdfsubject={cs},
pdfauthor={Andrew I.~Perminov},
pdfkeywords={Fast matrix multiplication, flip graph, serendipitous product, tensor rank},
}

\begin{document}
\maketitle

\begin{abstract}
This paper presents new results for fast matrix multiplication in small formats obtained by combining the meta flip graph framework with the serendipitous product construction. The framework has been extended to support all 680 rectangular formats with dimensions up to $16 \times 16 \times 16$. Compared to the previous state of the art, ranks are improved for \improvedCount\ formats. For \rediscoveredZT\ formats, ternary schemes are found where previously only integer or rational coefficients were known. Additionally, \betterStrassen\ new schemes with asymptotic exponent $\omega < \log_2 7$ are discovered, bringing the total number of such schemes to \betterStrassenTotal. The overall distribution of coefficient types across all investigated formats is \countZT\ ternary, \countZ\ integer, and \countQ\ rational. All code and discovered schemes are available as open source.
\end{abstract}


\input{structure/introduction}
\input{structure/preliminaries}
\input{structure/approach}
\input{structure/results}
\input{structure/discussion}
\input{structure/conclusion}

\section*{Acknowledgements}
The author would like to thank A. Sedoglavic for his valuable comments, criticism, and suggestions that improved the content of this paper, as well as for his explanations of the serendipitous product construction. Thanks are also due to Isaac Wood for explaining the principles of structural decomposition of schemes and how it can be used, for discussions of flip graph ideas, and for many fruitful conversations on the search for matrix multiplication schemes.

\bibliographystyle{unsrtnat}
\bibliography{references}

\end{document}

%% file: structure/introduction.tex
\section{Introduction}

Matrix multiplication is a fundamental operation in numerical linear algebra and beyond. The naive algorithm multiplies two $n \times n$ matrices using $n^3$ multiplications. Strassen's breakthrough~\cite{strassen1969gaussian} reduced the exponent to $\log_2 7 \approx 2.807$ by presenting a scheme for the $2 \times 2$ format, launching the search for Strassen-like algorithms. The rank of multiplying an $n \times m$ matrix by an $m \times p$ matrix is the minimum number $r$ for which an algorithm exists. Improvements in rank for small formats are of particular interest, as they can lead to better asymptotic exponents for matrix multiplication when combined through recursive or product constructions.

The best known ranks for all formats up to certain bounds are continuously tracked in an online catalog maintained by Sedoglavic~\cite{sedoglavic2025yet}, which currently contains 5426 entries and serves as the primary reference for upper bounds. Beyond the rank itself, the coefficient set used by a scheme is of practical importance. Schemes with coefficients in $\mathbb{Z}_T = \{-1,0,1\}$ (ternary coefficients) are particularly attractive for hardware implementation, as they require only additions, subtractions, and no multiplications by constants beyond sign changes. Schemes requiring larger integers ($\mathbb{Z}$) introduce multiplication by non-trivial constants, while schemes with fractions ($\mathbb{Q}$) require division (or shifts when dividing by powers of two) operations, both of which add computational overhead.

Various automated methods have been developed to discover matrix multiplication schemes. SAT solving with symmetry breaking~\cite{heule2019local, heule2021new, yang2024ruling} works for small formats but scales poorly. Constraint programming~\cite{deza2023fast} is effective for exhaustive search in small spaces. Numerical optimization~\cite{smirnov2013bilinear} often yields a lot of fractions and suffers from poor convergence to exact solutions. AlphaTensor~\cite{fawzi2022discovering} uses reinforcement learning and frames algorithm discovery as a game of tensor factorization. AlphaEvolve~\cite{novikov2025alphaevolve} extends this direction by employing language models to automate the search. However, these AI-discovered algorithms often employ large integers, fractions, or complex coefficients, introducing computational overhead for hardware implementation. More recent machine learning approaches train neural networks to directly discover low-rank decompositions~\cite{stapleton202560, elser2016network}.

An alternative paradigm that operates directly on algebraic structures is the flip graph approach. The flip graph approach~\cite{kauers2023flip} models schemes as vertices and local transformations as edges. The adaptive flip graph~\cite{arai2024adaptive} introduced the \texttt{plus} operator, which helps avoid getting stuck in local minima and accelerates the search for lower-rank decompositions. Subsequent work~\cite{moosbauer2025flip} incorporated symmetries of matrix multiplication into the flip graph framework, significantly reducing the search space and enabling more efficient exploration. The meta flip graph~\cite{kauers2025exploring} further extended this paradigm by introducing meta-operators that change the dimensions of schemes, enabling navigation between different matrix formats. The present work builds directly on the meta flip graph framework, continuing the direction established in previous work on flip graph search~\cite{perminov2025fast} and its open-source implementation~\cite{perminov2026fast}.

This paper presents new results obtained by combining the meta flip graph approach with the serendipitous product construction introduced by Smith~\cite{Smith:2002aa} and generalized by Sedoglavic~\cite{sedoglavic2025yet}. The main contributions are as follows:

\begin{itemize}
    \item Improved ranks for \improvedCount\ matrix multiplication formats compared to the previous state documented in~\cite{perminov2026fast}.
    \item Rediscovery of \rediscoveredZT\ schemes in $\mathbb{Z}_T$ (ternary coefficients) that were previously known only over $\mathbb{Z}$ or $\mathbb{Q}$.
    \item Discovery of \betterStrassen\ new schemes with asymptotic exponent $\omega < \log_2 7$, bringing the total number of such schemes to \betterStrassenTotal.
    \item A comprehensive survey of coefficient types across 680 formats $n \times m \times p$ with $2 \le n \le m \le p \le 16$: \countZT\ are ternary ($\mathbb{Z}_T$), \countZ\ require integers outside $\{-1,0,1\}$, and \countQ\ involve fractions.
\end{itemize}

All code and discovered schemes are available as open source at:
\begin{itemize}
    \item Flip graph framework: \url{https://github.com/dronperminov/ternary_flip_graph}
    \item Schemes and results: \url{https://github.com/dronperminov/FastMatrixMultiplication}
\end{itemize}

The paper is organized as follows. Section~\ref{sec:preliminaries} recalls the necessary definitions. Section~\ref{sec:approach} describes the meta flip graph framework as implemented and the procedure for exploiting serendipitous products. Section~\ref{sec:results} presents the experimental setup and the numerical results. Section~\ref{sec:discussion} discusses possible improvements and future directions. Section~\ref{sec:conclusion} concludes.

%% file: structure/preliminaries.tex
\section{Preliminaries}
\label{sec:preliminaries}

\subsection{Matrix Multiplication Schemes}

\begin{definition}[\textbf{Matrix multiplication scheme}]
A scheme for multiplying matrices $A \in \mathbb{F}^{n \times m}$ and $B \in \mathbb{F}^{m \times p}$ over an arbitrary field $\mathbb{F}$ with rank $r$ is denoted $(n,m,p:r)$. It consists of three tensors $U \in \mathbb{F}^{r \times n \times m}$, $V \in \mathbb{F}^{r \times m \times p}$ and $W \in \mathbb{F}^{r \times p \times n}$ that define the computation as follows.
\end{definition}

First, $r$ bilinear products (multiplications) are computed:

\begin{align*}
m_1 = (u^{(1)}_{11} a_{11} + \cdots + u^{(1)}_{nm} a_{nm}) \cdot & (v^{(1)}_{11} b_{11} + \cdots + v^{(1)}_{mp} b_{mp})\\
\vdots \\
m_r = (u^{(r)}_{11} a_{11} + \cdots + u^{(r)}_{nm} a_{nm}) \cdot & (v^{(r)}_{11} b_{11} + \cdots + v^{(r)}_{mp} b_{mp}),
\end{align*}

and the elements of the result matrix $C = AB$ are calculated as:

\begin{align*}
c_{ij} = w^{(1)}_{ji}m_1 + \cdots + w^{(r)}_{ji}m_r.
\end{align*}

For the scheme to be correct, the coefficients $u^{(l)}_{i_1 i_2}$, $v^{(l)}_{j_1 j_2}$, $w^{(l)}_{k_1 k_2}$ must satisfy a system of polynomial equations known as the Brent equations~\cite{brent1970algorithms}:

\begin{equation*}
\sum_{l=1}^{r} u^{(l)}_{i_1 i_2} v^{(l)}_{j_1 j_2} w^{(l)}_{k_1 k_2} = \delta_{i_2 j_1} \delta_{i_1 k_2} \delta_{j_2 k_1},
\end{equation*}

for all indices $i_1, k_2 \in \{1,\dots,n\}$, $i_2, j_1 \in \{1,\dots,m\}$, $j_2, k_1 \in \{1,\dots,p\}$. The Kronecker delta $\delta_{ij}$ equals $1$ if $i=j$ and $0$ otherwise.

Formally, each $u^{(l)}$ is an $n \times m$ matrix, $v^{(l)}$ is an $m \times p$ matrix, and $w^{(l)}$ is a $p \times n$ matrix. For notational convenience, these matrices are vectorized in row-major order and treated as vectors of dimensions $nm$, $mp$ and $pn$ respectively. Each such vectorized component is called a rank-one tensor. This vectorized view simplifies the algebraic manipulation of schemes and is used throughout the remainder of the paper.

\subsection{Flip Graph Operators}

The flip graph approach~\cite{kauers2023flip} represents each valid scheme as a vertex. Edges correspond to local transformations that preserve correctness. The core operators are defined as follows.

\begin{definition}[\textbf{Flip}]
For two rank-one tensors with $u^{(i)} = u^{(j)}$, the transformation is
\begin{align*}
    u^{(i)} \otimes v^{(i)} \otimes w^{(i)} \;+\; &u^{(j)} \otimes v^{(j)} \otimes w^{(j)} \;\rightarrow \\
    u^{(i)} \otimes (v^{(i)} + v^{(j)}) \otimes w^{(i)} \;+\; &u^{(j)} \otimes v^{(j)} \otimes (w^{(j)} - w^{(i)}).
\end{align*}
This preserves rank and correctness. The operator can be applied analogously when $v^{(i)} = v^{(j)}$ or $w^{(i)} = w^{(j)}$.
\end{definition}

\begin{definition}[\textbf{Plus}]
For two rank-one tensors satisfying $u^{(i)} \neq u^{(j)}$, $v^{(i)} \neq v^{(j)}$, and $w^{(i)} \neq w^{(j)}$, the transformation expands the scheme as:
\begin{align*}
    u^{(i)} \otimes v^{(i)} \otimes w^{(i)} \quad+&\quad u^{(j)} \otimes v^{(j)} \otimes w^{(j)} \quad \rightarrow\\
u^{(i)} \otimes (v^{(i)} + v^{(j)}) \otimes w^{(i)} \quad+&\quad u^{(i)} \otimes v^{(j)} \otimes (w^{(j)} - w^{(i)}) \quad+\quad (u^{(j)} - u^{(i)}) \otimes v^{(j)} \otimes w^{(j)}
\end{align*}
This operation increases the scheme's rank while preserving correctness.
\end{definition}

\begin{definition}[\textbf{Reduction}]
For two rank-one tensors with $u^{(i)} = u^{(j)}$ and $v^{(i)} = v^{(j)}$, the transformation combines them as:
    \begin{align*}
        u^{(i)} \otimes v^{(i)} \otimes w^{(i)} \;+\; u^{(j)} \otimes v^{(j)} \otimes w^{(j)} \;\rightarrow\; u^{(i)} \otimes v^{(i)} \otimes (w^{(i)} + w^{(j)}).
    \end{align*}
    This decreases the rank by one by eliminating redundant components. The operator applies analogously for equalities in other pairs of tensors.
\end{definition}

The meta flip graph framework~\cite{kauers2025exploring} introduces operators that change dimensions, enabling navigation between different matrix formats:

\begin{definition}[\textbf{Merge}]
Combines two schemes with compatible dimensions. For example, $(n,m,p_1: r_1)$ and $(n,m,p_2: r_2)$ merge to $(n,m,p_1+p_2: r_1+r_2)$. The same applies to the other two axes.
\end{definition}
    
\begin{definition}[\textbf{Product}]
Forms the tensor product of two schemes. Given $(n_1,m_1,p_1: r_1)$ and $(n_2,m_2,p_2: r_2)$, the result is $(n_1 n_2, m_1 m_2, p_1 p_2: r_1 r_2)$.
\end{definition}

\begin{definition}[\textbf{Extend}]
Naively extends a scheme by one dimension. For instance, $(n,m,p: r)$ extends to $(n,m,p+1: r + m \cdot n)$ by merging with a trivial $(n,m,1: m \cdot n)$ scheme that computes each output entry directly. Analogous extensions exist for the other axes.
\end{definition}
    
\begin{definition}[\textbf{Project}]
Reduces a scheme by one dimension. From $(n,m,p: r)$, projection along the $p$-axis yields $(n,m,p-1: r')$ where $r' \le r$, obtained by removing entries corresponding to the last dimension and eliminating zero coefficients. The other axes are handled similarly.
\end{definition}

\subsection{Coefficient Rings}

Throughout this paper, the following coefficient rings are used.

\begin{notation}[$\mathbb{Z}_2$]
Binary coefficients $\{0,1\}$ with arithmetic modulo $2$.
\end{notation}

\begin{notation}[$\mathbb{Z}_3$]
Ternary coefficients $\{0,1,2\}$ with arithmetic modulo $3$.
\end{notation}

\begin{notation}[$\mathbb{Z}_T$]
Integer ternary coefficients $\{-1,0,1\}$ with standard integer arithmetic.
\end{notation}

\begin{notation}[$\mathbb{Z}$]
Integer coefficients, where at least one coefficient satisfies $|c| > 1$.
\end{notation}

\begin{notation}[$\mathbb{Q}$]
Rational coefficients, where at least one coefficient is a non-integer fraction.
\end{notation}

\subsection{Tensor Structure of Matrix Multiplication Schemes}
\label{subsec:tensor_structure}

A matrix multiplication scheme $(n,m,p: r)$ can be represented as a sum of rank-one tensors:

\begin{equation}
\mathcal{T} = \sum_{i=1}^{r} u^{(i)} \otimes v^{(i)} \otimes w^{(i)}.
\end{equation}

\begin{definition}[\textbf{Bud}\footnote{This terminology was suggested by A. Sedoglavic in private communication.}]
A pair of indices $(i,j)$, $i \neq j$, such that $u^{(i)} = u^{(j)}$ (or the same condition holds for $v$ or $w$). If equality holds for more than one tensor among $\{U,V,W\}$, the rank can be reduced, therefore only buds where equality holds for exactly one tensor are considered.
\end{definition}

\begin{definition}[\textbf{Elementary matrix multiplication tensor}]
A set of indices together with the corresponding components $u^{(i)}$, $v^{(i)}$, $w^{(i)}$ that form a self-contained matrix multiplication substructure of dimensions $N \times M \times P$, denoted $\langle N, M, P \rangle$.
\end{definition}

A scheme can be decomposed into elementary tensors formed by its buds~\cite{kauers2026exploiting}. The buds induce relations on the set of indices: equalities in $U$, $V$ and $W$ collectively suggest possible groupings. Indices can be grouped into blocks, where each block corresponds to an elementary matrix multiplication tensor $\langle N_i, M_i, P_i \rangle$, with $N_i$, $M_i$, $P_i$ determined by the bud pattern within that block. The remaining indices (those not participating in any block) form trivial blocks $\langle 1,1,1 \rangle$.

\begin{example}[\textbf{Elementary tensors from bud patterns}]~

\begin{itemize}
    \item A single index: $\langle 1,1,1 \rangle$.
    \item A $U$-bud $(i,j)$: $\langle 1,1,2 \rangle$.
    \item $k$ indices with all pairwise $U$-buds: $\langle 1,1,k \rangle$.
    \item A $V$-bud $(i,j)$: $\langle 2,1,1 \rangle$.
    \item A $W$-bud $(i,j)$: $\langle 1,2,1 \rangle$.
    \item Four indices $\{i_1,i_2,j_1,j_2\}$ with $U$-buds $(i_1,i_2)$, $(j_1,j_2)$ and $V$-buds $(i_1,j_1)$, $(i_2,j_2)$: $\langle 2,1,2 \rangle$.
\end{itemize}
\end{example}

The same set of buds can sometimes be grouped into elementary tensors in different ways. For example, a $\langle 2,1,2 \rangle$ tensor can also be represented as two independent $\langle 1,1,2 \rangle$ or two independent $\langle 2,1,1 \rangle$ tensors.

\begin{definition}[\textbf{Scheme tensor structure}]
A representation of a scheme as a sum of elementary tensors:
\begin{equation}
\mathcal{T} = \sum_{i=1}^{t} S_i \cdot \langle N_i, M_i, P_i \rangle,
\end{equation}
where each $\langle N_i, M_i, P_i \rangle$ is an elementary matrix multiplication tensor formed by a chosen grouping of indices, and $S_i$ is a multiplicity factor.
\end{definition}

The rank of the scheme then satisfies

\begin{equation}
r = \sum_{i=1}^{t} S_i \cdot N_i \cdot M_i \cdot P_i.
\end{equation}

For a scheme without any buds, the decomposition consists of $r$ different $\langle 1,1,1 \rangle$ tensors.

\subsection{Serendipitous Product}
\label{subsec:serendipitous_product}

The serendipitous product construction, introduced by Smith~\cite{Smith:2002aa} and generalized by Sedoglavic~\cite{sedoglavic2025yet}, combines two matrix multiplication schemes $(n_1, m_1, p_1: r_1)$ and $(n_2, m_2, p_2: r_2)$ to obtain a scheme for the format $(n_1 n_2, m_1 m_2, p_1 p_2: r_s)$ whose rank can be lower than the naive (Kronecker) product $r_1 \cdot r_2$.

Let the first scheme be decomposed into elementary tensors:
\begin{equation}
\mathcal{T}_1 = \sum_{i=1}^{t} S_i \cdot \langle N_i, M_i, P_i \rangle.
\end{equation}

\begin{definition}[\textbf{Serendipitous product}]
Applying the second scheme to each elementary blocks gives
\begin{equation}
\mathcal{T}_1 \otimes (n_2, m_2, p_2: r_2) \;=\; \sum_{i=1}^{t} S_i \cdot \langle N_i n_2,\; M_i m_2,\; P_i p_2 \rangle.
\end{equation}

Each term $\langle N_i n_2, M_i m_2, P_i p_2 \rangle$ is realized using the best known scheme for that format, with rank $R(N_i n_2, M_i m_2, P_i p_2)$. The concrete algorithm for constructing such a product is described in Section~\ref{subsec:serendipitous_product_construction}.
\end{definition}

The total rank of the serendipitous product is therefore

\begin{equation}
\label{eq:serendipitous_rank}
r_s = \sum_{i=1}^{t} S_i \cdot R\bigl(N_i n_2,\; M_i m_2,\; P_i p_2\bigr),
\end{equation}

where $R(N,M,P)$ denotes the best known rank for format $(N,M,P)$.

The construction yields a genuine improvement over the naive product only when $R(N_i n_2, M_i m_2, P_i p_2) < N_i M_i P_i \cdot R(n_2, m_2, p_2)$ for at least one block. For blocks where equality holds, using the classical Kronecker product with the second scheme directly is simpler, as no saving is achieved.

In the naive case, the first scheme is represented trivially as $r_1$ copies of $\langle 1,1,1 \rangle$, i.e., $\mathcal{T}_1 = r_1 \cdot \langle 1,1,1 \rangle$. Then the construction reduces to the Kronecker tensor product, yielding $\langle n_2, m_2, p_2 \rangle$ repeated $r_1$ times, which corresponds exactly to the naive scheme of rank $r_1 \cdot r_2$. The formula above correctly reproduces this as $r_n = r_1 \cdot r_2$.

\subsection{Construction of Serendipitous Product}
\label{subsec:serendipitous_product_construction}

Let the first scheme be $(n_1, m_1, p_1: r_1)$ with tensors $U_1$, $V_1$, $W_1$, and the second scheme $(n_2, m_2, p_2: r_2)$ with tensors $U_2$, $V_2$, $W_2$. The indices $\{1,\dots,r_1\}$ are partitioned into blocks according to the elementary tensor decomposition $\sum_i^t S_i \langle N_i, M_i, P_i \rangle$. The result of the serendipitous product is a scheme of dimensions $(n_1 n_2, m_1 m_2, p_1 p_2)$ whose rank $r_s$ is the sum over blocks of $R(N_i n_2, M_i m_2, P_i p_2)$.

This section describes how to construct the tensors $U$, $V$, $W$ of the resulting scheme for each elementary block. The construction is presented for the basic cases, more complex blocks are handled by combining these rules.

\subsubsection{Trivial block $\langle 1,1,1 \rangle$}

A single index $i$ in the first scheme contributes $r_2$ rank-one tensors via the Kronecker product:
\begin{align*}
u^{(j)} &= u_1^{(i)} \otimes u_2^{(j)}, \\
v^{(j)} &= v_1^{(i)} \otimes v_2^{(j)}, \\
w^{(j)} &= w_1^{(i)} \otimes w_2^{(j)},
\end{align*}
for $j = 1,\dots,r_2$.

\begin{figure}[ht!]
\centering
\includegraphics[width=\textwidth]{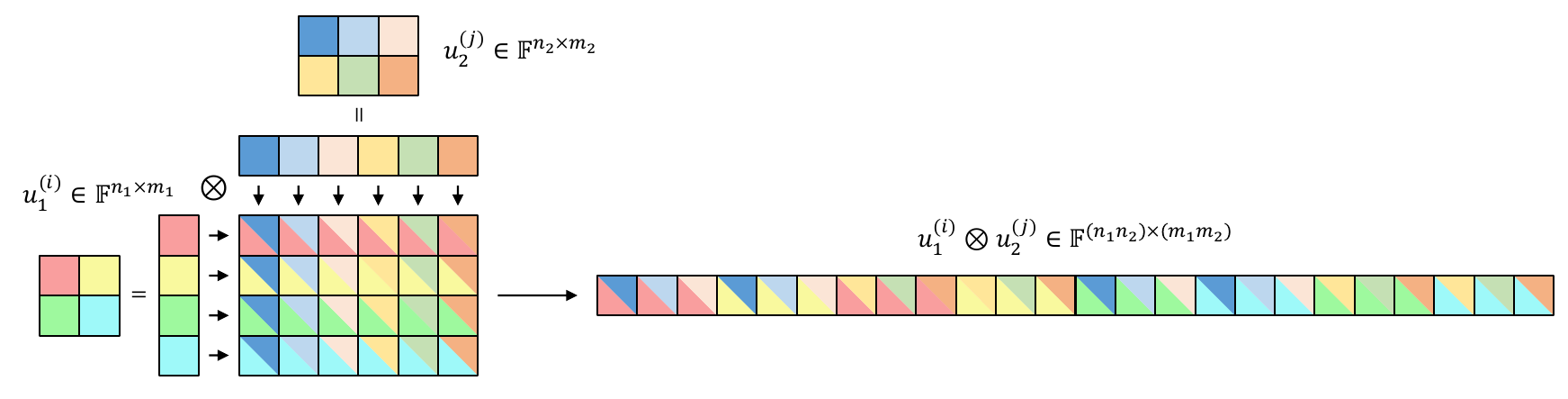}
\caption{Kronecker product of a single $n_1 \times m_1$ row $u_1^{(i)}$ from the first scheme with a single $n_2 \times m_2$ row $u_2^{(j)}$ from the second scheme. The result is a single row of size $(n_1 n_2) \times (m_1 m_2)$ for the $U$ tensor of the resulting scheme.}
\label{fig:product_1_by_1}
\end{figure}

Figure~\ref{fig:product_1_by_1} illustrates this product: a $2 \times 2$ row multiplied by a $2 \times 3$ row produces a $4 \times 6$ row. The figure shows the general case with dimensions $n_1$, $m_1$, $n_2$, $m_2$.

\begin{figure}[ht!]
\centering
\includegraphics[width=\textwidth]{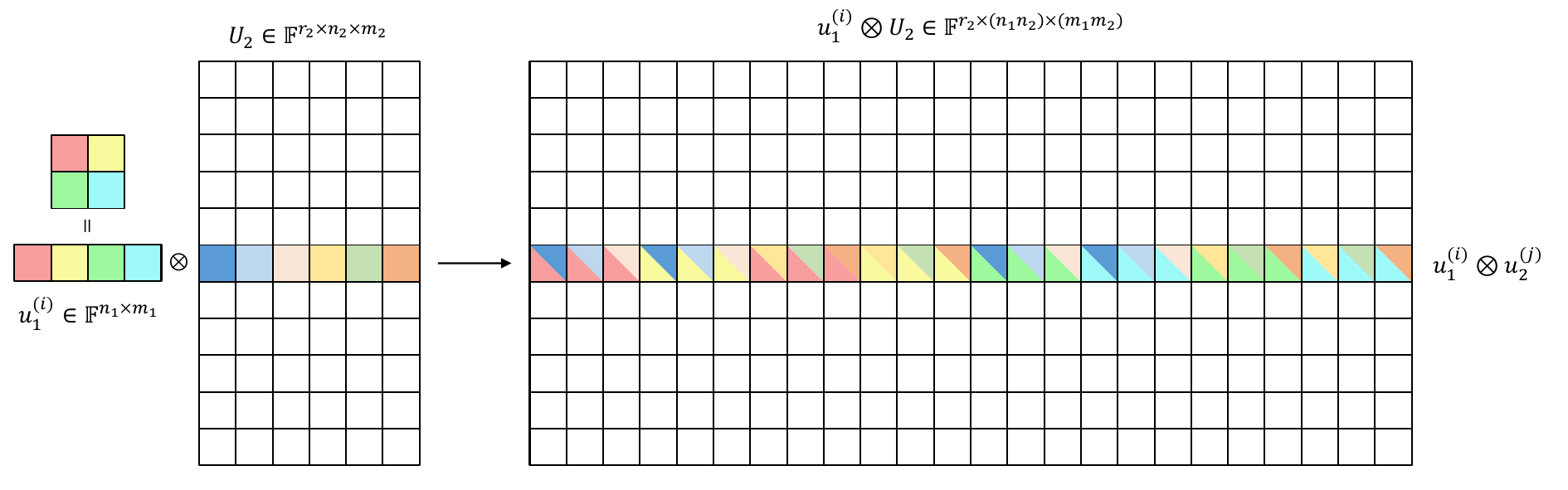}
\caption{Kronecker product of a single $u_1^{(i)}$ from the first scheme with all $r_2$ rows $u_2^{(1)},\dots,u_2^{(r_2)}$ from the second scheme. The result is $r_2$ rows for the $U$ tensor of the resulting scheme, each of size $(n_1 n_2) \times (m_1 m_2)$.}
\label{fig:product_1_by_r}
\end{figure}

Repeating this for all $r_2$ rows of the second scheme yields the full contribution of the trivial block, as shown in Figure~\ref{fig:product_1_by_r}. There, a single index is multiplied by an $11 \times 2 \times 3$ tensor, producing an $11 \times 4 \times 6$ tensor.

\subsubsection{Bud block in $U$: $\langle 1,1,k \rangle$}

Now consider a bud group in $U$ consisting of $k$ indices $i_1, i_2, \dots, i_k$ with $u_1^{(i_1)} = u_1^{(i_2)} = \cdots = u_1^{(i_k)}$. Instead of forming $k \cdot r_2$ triplets naively, a scheme for $(n_2, m_2, k p_2: r_2')$ (only if $r_2' < k\cdot r_2$) is used, with tensors $U_3$, $V_3$, $W_3$. The resulting $r_2'$ triplets are constructed as follows.

For the $U$ tensor, the construction is identical to the trivial block using the common $u_1^{(i_1)}$-vector:
\begin{equation*}
u^{(j)} = u_1^{(i_1)} \otimes u_3^{(j)}.
\end{equation*}

For the $V$ tensor, each entry is formed by summing contributions from all $k$ indices. The third scheme is split into $k$ consecutive blocks along its $p$ dimension, each block of size $m_2 \times p_2$. For a fixed index $l$ from $1$ to $k$, the contribution uses the $l$-th block of the third scheme's $V_3$ tensor. The sum runs over $l$, multiplying the $v_1^{(i_l)}$ component from the first scheme by the corresponding entry from the $l$-th block of $V_3$.

For the $W$ tensor, the construction is analogous, but the splitting is applied to the other dimension of the third scheme. The $W_3$ tensor is split into $k$ consecutive blocks along its first dimension (size $p_2 \times n_2$ per block). The $l$-th contribution uses the $l$-th block of $W_3$, multiplied by $w_1^{(i_l)}$ from the first scheme.

\begin{figure}[ht!]
\centering
\includegraphics[width=\textwidth]{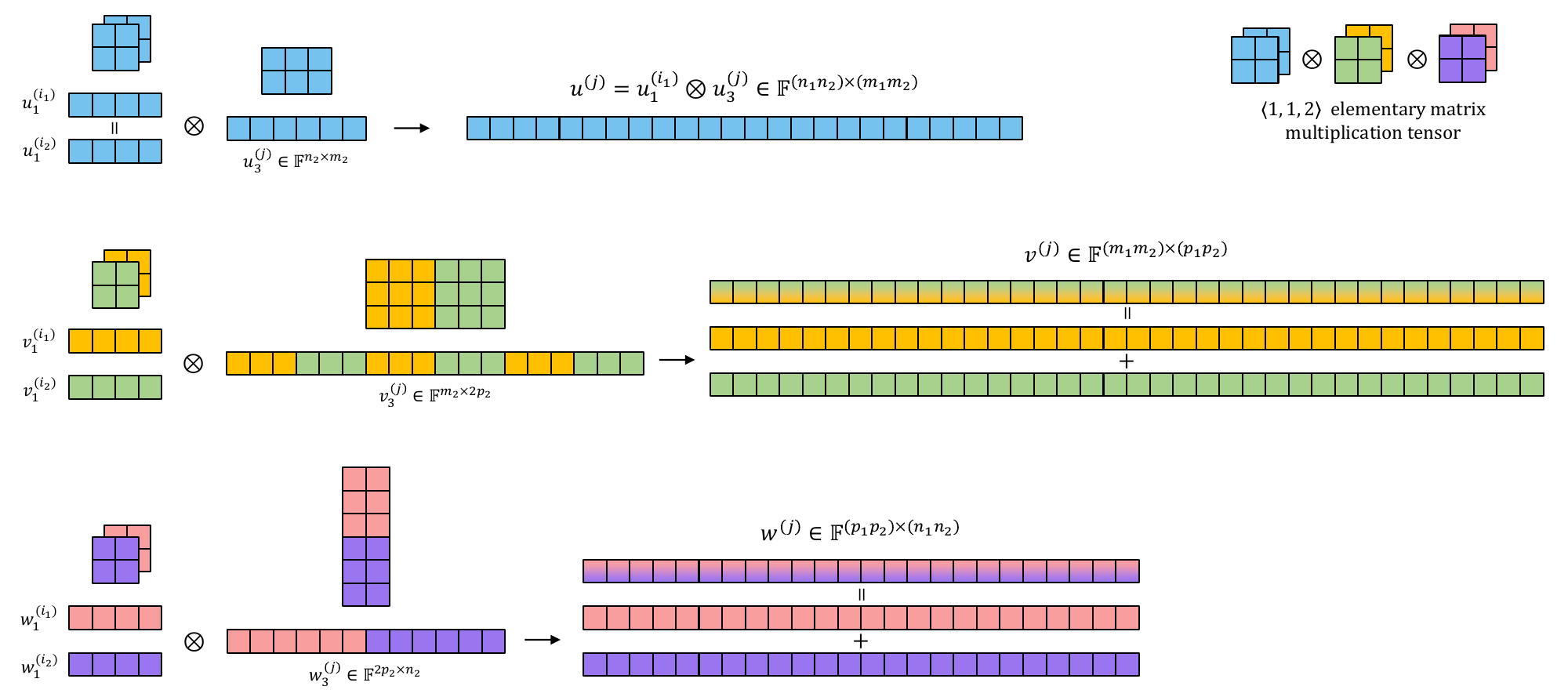}
\caption{Serendipitous product for a $\langle 1,1,2 \rangle$ block multiplied by one row of the third scheme $(n_2, m_2, 2p_2: r_2')$. The common $u$-vector yields one row of $U$; $V$ and $W$ sum contributions from both indices across the two blocks.}
\label{fig:product_112}
\end{figure}

Thus, the third scheme is effectively divided into $k$ blocks along the $p$ dimension, each assigned to one of the $k$ indices. Figure~\ref{fig:product_112} illustrates this construction for $k=2$: a $\langle 1,1,2 \rangle$ tensor multiplied by a single row of the third scheme $(n_2, m_2, 2p_2: r_2')$ produces a single row of the resulting $U$, $V$, and $W$ tensors. The figure shows how the common $u$-vector generates the corresponding row of the $U$ tensor, while $V$ and $W$ sum contributions from both indices across the two blocks. Repeating this process for all $r_2'$ rows of the third scheme yields the full tensors.

\subsubsection{Bud blocks in $V$ or $W$}

The cases $\langle k,1,1 \rangle$ (buds in $V$) and $\langle 1,k,1 \rangle$ (buds in $W$) are handled similarly, with the splitting applied to the appropriate dimension.

For $\langle k,1,1 \rangle$, take $k$ indices sharing the same $v$-vector. Use a scheme for $(k n_2, m_2, p_2: r_2')$. The common $v$ forms $V$ via Kronecker product. The third scheme is split into $k$ blocks along the $n$ dimension, and $U$ and $W$ sum the contributions from each block.

For $\langle 1,k,1 \rangle$, take $k$ indices sharing the same $w$-vector. Use a scheme for $(n_2, k m_2, p_2: r_2')$. The common $w$ forms $W$ via Kronecker product. The third scheme is split into $k$ blocks along the $m$ dimension, and $U$ and $V$ sum the contributions from each block.

\subsubsection{Blocks with multiple non-unit dimensions}

For elementary tensors where more than one of $N_i, M_i, P_i$ exceeds $1$, e.g., $\langle 2,1,2 \rangle$, the construction combines the rules for the respective axes. The third scheme is split into blocks along the corresponding dimensions, and the resulting tensors $U$, $V$, $W$ are formed by summing over the appropriate Cartesian products of indices. The same principle applies to all elementary tensors arising from bud patterns.

\subsubsection{Combining the Blocks}
After processing each elementary block $\langle N_i, M_i, P_i \rangle$ as described above, the resulting rank-one tensors from all blocks are concatenated to form the final scheme. The total rank $r_s$ is therefore given by Equation~\ref{eq:serendipitous_rank}.

The $U$, $V$ and $W$ tensors of the resulting scheme are obtained by stacking the corresponding tensors from each block along the rank dimension.

\begin{example}[\textbf{Serendipitous product for $(2,3,4:20)$ and $(2,4,2:14)$}]~

Consider the scheme $(2,3,4:20)$ with the following tensors $U_1$, $V_1$, $W_1$:

\begin{scriptsize}
\setlength{\arraycolsep}{1.7pt}
\[
U_1 = \left[\begin{array}{rrrrrr}
    \rowcolor{green!20} 0 & 0 & 0 & 0 & 1 & -1 \\
    \rowcolor{yellow!20} 0 & 0 & 0 & 1 & 0 & 0 \\
    \rowcolor{red!15} 0 & 1 & -1 & 0 & 0 & 0 \\
    0 & 0 & -1 & 0 & 0 & 0 \\
    0 & 0 & -1 & 0 & 0 & -1 \\
    \rowcolor{blue!15} -1 & 0 & 0 & 0 & 0 & 0 \\
    \rowcolor{red!15} 0 & -1 & 1 & 0 & 0 & 0 \\
    0 & 0 & 1 & 1 & 0 & 1 \\
    0 & 0 & 1 & 0 & 1 & 0 \\
    -1 & -1 & 0 & 0 & 0 & 0 \\
    \rowcolor{yellow!20} 0 & 0 & 0 & 1 & 0 & 0 \\
    -1 & 0 & -1 & -1 & 0 & -1 \\
    0 & -1 & 0 & 0 & -1 & 0 \\
    1 & 1 & 0 & 0 & 1 & 0 \\
    \rowcolor{green!20} 0 & 0 & 0 & 0 & 1 & -1 \\
    -1 & -1 & 0 & -1 & 0 & -1 \\
    \rowcolor{blue!15} -1 & 0 & 0 & 0 & 0 & 0 \\
    0 & 0 & 0 & 0 & 1 & 0 \\
    0 & 0 & 0 & 1 & 0 & 1 \\
    -1 & -1 & 0 & -1 & -1 & 0
\end{array}\right]\quad
V_1 = \left[\begin{array}{rrrrrrrrrrrr}
    0 & 0 & 0 & 0 & 0 & 0 & 0 & 0 & 0 & 0 & 0 & -1 \\
    -1 & 0 & 0 & 0 & 1 & 0 & 0 & 0 & 1 & 0 & 0 & 0 \\
    0 & 0 & 0 & 0 & 0 & 0 & 0 & 0 & 0 & -1 & 0 & 0 \\
    0 & 0 & 0 & 0 & 0 & 0 & -1 & 0 & 0 & 0 & -1 & 0 \\
    1 & 0 & -1 & 0 & -1 & 0 & 0 & 0 & -1 & 0 & 1 & -1 \\
    0 & 0 & -1 & 0 & 0 & 0 & 0 & 0 & 0 & 0 & 0 & 0 \\
    0 & 0 & 0 & 0 & 0 & 0 & -1 & 0 & 0 & 1 & 0 & 0 \\
    -1 & 0 & 1 & 0 & 1 & 0 & 0 & 0 & 1 & 0 & 0 & 0 \\
    0 & 0 & 0 & 0 & 0 & 0 & -1 & 0 & 0 & 0 & 0 & -1 \\
    0 & 0 & 0 & 0 & 0 & -1 & 0 & 0 & 0 & -1 & 0 & 0 \\
    0 & 0 & 0 & -1 & 0 & 0 & 0 & 0 & 0 & 0 & 0 & 0 \\
    1 & 0 & -1 & 0 & -1 & 0 & 0 & 0 & 0 & 1 & 0 & 0 \\
    0 & 1 & 0 & -1 & 0 & -1 & -1 & 1 & 0 & -1 & 0 & 0 \\
    0 & -1 & 0 & 1 & 0 & 1 & 0 & 0 & 0 & 1 & 0 & 0 \\
    0 & 0 & 0 & 0 & 1 & 0 & 0 & 0 & 0 & 0 & 0 & 1 \\
    0 & 0 & 0 & 0 & 1 & 0 & 0 & 0 & 0 & -1 & 0 & 0 \\
    0 & 1 & -1 & 0 & 0 & -1 & 0 & 0 & 0 & -1 & 0 & 0 \\
    0 & 0 & 0 & 0 & 0 & 0 & 0 & -1 & 0 & 0 & 0 & -1 \\
    0 & 0 & 0 & 0 & 1 & 0 & 0 & 0 & 1 & 0 & 0 & 0 \\
    0 & 1 & 0 & -1 & 1 & 0 & 0 & 0 & 0 & -1 & 0 & 0
\end{array}\right]\quad
W_1 = \left[\begin{array}{rrrrrrrr}
    0 & 1 & 0 & -1 & 0 & 1 & 0 & 1 \\
    0 & -1 & 0 & 0 & 0 & -1 & 0 & 0 \\
    -1 & 0 & 1 & 0 & -1 & 0 & -1 & 0 \\
    0 & 0 & 0 & 0 & 1 & -1 & 0 & 0 \\
    0 & 0 & 0 & 0 & 0 & -1 & 0 & 0 \\
    1 & 0 & 1 & 0 & 1 & 0 & 1 & 0 \\
    0 & 0 & 0 & 0 & 1 & 0 & 1 & 0 \\
    1 & 0 & 0 & 0 & 0 & 1 & 0 & 0 \\
    0 & 0 & 0 & 0 & 0 & -1 & -1 & 0 \\
    0 & 0 & 1 & -1 & 0 & 0 & 0 & 0 \\
    0 & 0 & 0 & -1 & 0 & 0 & 0 & -1 \\
    -1 & 0 & 0 & 0 & 0 & 0 & 0 & 0 \\
    0 & 0 & 0 & 0 & 0 & 0 & -1 & 0 \\
    0 & 0 & 0 & 1 & 0 & 0 & 1 & 0 \\
    0 & 1 & 0 & -1 & 0 & 0 & 0 & 0 \\
    -1 & 0 & 0 & 1 & 0 & 0 & 0 & 0 \\
    0 & 0 & -1 & 0 & 0 & 0 & -1 & 0 \\
    0 & 0 & 0 & 0 & 0 & 0 & 1 & -1 \\
    -1 & 1 & 0 & 0 & 0 & 0 & 0 & 0 \\
    0 & 0 & 0 & -1 & 0 & 0 & 0 & 0
\end{array}\right]
\]
\end{scriptsize}

This scheme has four disjoint $U$-buds: $(1,15)$, $(2,11)$, $(3,7)$, $(6,17)$. Each bud forms a $\langle 1,1,2 \rangle$ tensor. The remaining $12$ indices form trivial $\langle 1,1,1 \rangle$ tensors. Thus the decomposition is $4\langle 1,1,2 \rangle + 12\langle 1,1,1 \rangle$.

The naive Kronecker product with $(2,4,2:14)$ would give rank $20 \cdot 14 = 280$ for the format $(4,12,8)$. Using the serendipitous product,
\begin{equation*}
r_s = 4 \cdot R(2,4,4) + 12 \cdot R(2,4,2) = 4 \cdot 26 + 12 \cdot 14 = 272,
\end{equation*}
since $R(2,4,4) = 26 < 2 \cdot 14 = 28$. The saving of $8$ multiplications comes from the four $\langle 1,1,2 \rangle$ blocks.

The construction proceeds as follows:
\begin{itemize}
    \item For the $12$ trivial indices $\{4, 5, 8, 9, 10, 12, 13, 14, 16, 18, 19, 20\}$, form Kronecker products with $(2,4,2:14)$ directly. Each such index contributes $14$ triplets, yielding $12 \cdot 14 = 168$ triplets in total.
    \item For the bud $(1,15)$ (indices sharing the same $u$-vector), use the scheme $(2,4,4:26)$. It contributes $26$ triplets instead of $2 \cdot 14 = 28$ from the naive product. The third scheme is split into two blocks along the $p$ dimension: the first block corresponds to index $1$, the second to index $15$.
    \item For the bud $(2,11)$, similarly use $(2,4,4:26)$, contributing another $26$ triplets.
    \item For the bud $(3,7)$, use $(2,4,4:26)$, contributing $26$ triplets.
    \item For the bud $(6,17)$, use $(2,4,4:26)$, contributing $26$ triplets.
\end{itemize}
Concatenating all triplets gives the resulting scheme $(4,12,8:272)$, since $168 + 4 \cdot 26 = 272$.
\end{example}

%% file: structure/approach.tex
\section{Approach}
\label{sec:approach}

\subsection{Meta Flip Graph Framework}

The search is organized as a parallel exploration of the space of matrix multiplication schemes across different formats and ranks. The framework maintains a collection of schemes organized in a three-level hierarchy:

\begin{itemize}
    \item A \textbf{scheme store} holds schemes $(n,m,p: r)$ for a fixed format $n\times m\times p$ and rank $r$.
    \item A \textbf{rank store} for a given format $n\times m\times p$ maps ranks $r$ to scheme stores.
    \item A \textbf{format store} maps format triples $(n, m, p)$ to rank stores.
\end{itemize}

When a new scheme is discovered, a hash is computed based on the sorted list of its component strings $(u^{(i)}, v^{(i)}, w^{(i)})$. This prevents duplicate insertion of schemes that differ only by a permutation of the $r$ components. Equivalence under De Groote orbits is not checked, as this would be computationally prohibitive.

The search starts from an initial non-empty set of schemes. These can be naive schemes (e.g., $(2,2,2:8)$, $(2,2,3:12)$, $(2,3,3:18)$), best-known schemes obtained from the literature, or any other available schemes. For each such scheme, the corresponding format, rank and scheme stores are created if they do not already exist.

The algorithm runs $N$ independent workers in parallel. Each worker repeatedly performs the following steps:

\begin{enumerate}
    \item Select a format $n \times m \times p$ with probability inversely proportional to the number of minimal-rank schemes currently available for that format. Formats with fewer such schemes are more likely to be selected, encouraging exploration where progress is most needed.
    
    \item Select a rank $r$ for the chosen format using exponential weights: the smallest available rank $r_{\min}$ receives weight $1$, the next $r_{\min}+1$ receives weight $\alpha$ (with $0 < \alpha <= 1$) and so on. Typical values used are $\alpha = 0.5$ or $\alpha = 0.7$. This biases the search toward lower ranks.

    \item Select a scheme uniformly at random from the scheme store for $(n,m,p:r)$, or with probability proportional to the number of buds available in the scheme.

    \item Perform a random walk in the flip graph starting from this scheme. At each step, a flip is applied whenever possible. Occasionally, or when no flip is available, a plus operator is applied instead, increasing the rank and allowing the walk to escape local minima.

    \item If a scheme with rank $r-1$ is discovered during the walk, it is added to the collection. Then, meta-operators are applied to this new scheme: \texttt{project}, \texttt{extend}, \texttt{merge} and \texttt{product}. Each resulting scheme is added to the collection only if its rank does not exceed the best known rank for that format by more than a predefined threshold (typically 5–10). This prevents explosion of high-rank schemes.

    Additionally, if the worker is still using a scheme of rank $r$ (the same rank as the starting scheme) and a certain number of steps have been performed, with some probability this scheme is also added to the store as an alternative scheme for the same rank, and meta-operators are applied to it. This helps increase diversity in the collection without requiring a rank reduction.

    The worker then selects a new format and rank (according to the same probabilistic rules) and repeats.

    \item If the worker has performed a certain number of iterations without finding any rank reduction, it terminates the current random walk and returns to step 1 to select a new scheme. Typical thresholds used in the experiments ranged from 5 to 20 million steps without improvement.
\end{enumerate}

\subsection{Search in Different Coefficient Rings}

The framework supports three coefficient rings: $\mathbb{Z}_2$, $\mathbb{Z}_3$ and $\mathbb{Z}_T$. Search in $\mathbb{Z}_3$ was rarely used because rank reductions in $\mathbb{Z}_3$ are significantly slower to obtain compared to $\mathbb{Z}_2$ and $\mathbb{Z}_T$, making it less practical despite potential theoretical interest.

\textbf{Search in $\mathbb{Z}_2$ with Hensel lifting.} Direct search in $\mathbb{Z}_2$ produces schemes that often cannot be lifted to $\mathbb{Z}$, $\mathbb{Z}_T$ or $\mathbb{Q}$ (e.g., $(3,3,8:55)$, $(4,4,4:47)$, $(4,4,5:60)$, $(4,5,5:73)$). Such schemes, when propagated via meta-operators, lead to large numbers of degenerate schemes that waste computational resources. To mitigate this, each newly found $\mathbb{Z}_2$ scheme is subjected to Hensel lifting (up to 50 steps). Lifting continues until either it fails or rational reconstruction produces a valid scheme over $\mathbb{Z}$ or $\mathbb{Q}$ (verified via Brent equations). Only schemes that pass this check are stored and used in meta-operations. This reduces the presence of non-liftable degenerate schemes while preserving the speed of $\mathbb{Z}_2$ search.

\textbf{Search in $\mathbb{Z}_T$ (direct).} For larger formats (specifically those where the number of matrix entries exceeds 100, i.e., $n \cdot m \ge 100$ or $m \cdot p \ge 100$ or $p \cdot n \ge 100$), Hensel lifting becomes too slow. Therefore, the search runs directly in $\mathbb{Z}_T$ using the same meta flip graph procedure. This avoids lifting entirely and works for all formats up to $16 \times 16 \times 16$.

\subsection{Exploiting Serendipitous Products}

Each scheme $(n_1, m_1, p_1: r_1)$ in the collection is examined for its elementary tensor decomposition (see Section~\ref{subsec:tensor_structure}). Different possible decompositions are considered by choosing how to group indices into elementary tensors $\langle N_i, M_i, P_i \rangle$. A greedy random selection is repeated multiple times (typically 30 iterations) to obtain different candidate decompositions.

For each candidate decomposition, the serendipitous product construction is applied with all possible dimensions $(n_2, m_2, p_2)$ such that the resulting dimensions $(n, m, p) = (n_1 n_2,\; m_1 m_2,\; p_1 p_2)$ do not exceed $16 \times 16 \times 16$. The total rank of the serendipitous product ($r_s$) is then computed according to the formula in Section~\ref{subsec:serendipitous_product}. The combination (base scheme and its structure) with the smallest resulting rank is retained. If this rank improves the known rank for the target format, a new scheme is recorded.

For schemes $(n_1, m_1, p_1; r_1)$ where an improvement is found, an additional refinement step is performed: a short random walk in the flip graph (100 to 10000 steps) is applied, and after each step the buds are re-analyzed and serendipitous products are re-evaluated. This often leads to further rank improvements beyond the initial discovery. The resulting product schemes are then added to the collection.

Several examples of such constructions obtained in this work are summarized in Table~\ref{tab:serendipitous_examples} (the dimensions $(n, m, p)$ are presented in non-decreasing order, as permuting dimensions does not affect the rank). It is worth noting that the schemes used for the serendipitous product are not always optimal in rank. In many cases, a scheme with a rank higher than the best known (by 1, 2, 5 or even more) may yield better overall savings due to a richer bud structure. For instance, the format $(3,4,13)$ has an optimal rank of 117, yet a scheme of rank 123 was used successfully in a serendipitous construction.

\begin{table}[ht!]
	\caption{Examples of serendipitous product constructions. The naive rank $r_n$ is $r_1 \cdot r_2$.}
    \label{tab:serendipitous_examples}
	\centering
	\begin{tabular}{cccccc}
        \toprule
        First scheme & \multirow{2}{*}{First scheme structure} & Second scheme & \multirow{2}{*}{$n \times m \times p$} & \multicolumn{2}{c}{Product rank} \\
        $(n_1, m_1, p_1: r_1)$ & & $(n_2, m_2, p_2: r_2)$ & & $r_s$ & $r_n$ \\
        \midrule
$(2, 2, 7: 25)$ & $3\langle1, 1, 3\rangle + 4\langle1, 1, 4\rangle$ & $(7, 8, 1: 56)$ & $7\times14\times16$ & 1022 & 1400 \\
\addlinespace[2ex]
\multirow{2}{*}{$(2,3,8: 40)$} & \multirow{2}{*}{$4\langle1,1,4\rangle + 12\langle1,1,2\rangle$} & $(4,5,1: 20)$ & $8 \times 8 \times 15$ & 628 & 800 \\ 
 &  & $(4,4,2: 26)$ & $8 \times 12 \times 16$ & 960 & 1040 \\
\addlinespace[2ex]
\multirow{3}{*}{$(2, 3, 14: 70)$} & \multirow{3}{*}{$2\langle1, 1, 3\rangle+6\langle1, 1, 4\rangle+8\langle1, 1, 5\rangle$} & $(7,5,1:35)$ & $14 \times14 \times15$ & 1798 & 2450 \\
& & $(2, 3, 1: 6)$ & $4 \times 9 \times 14$ & 350 & 420 \\
& & $(2, 4, 1: 8)$ & $4 \times 12 \times 14$ & 452 & 560 \\
\addlinespace[2ex]
$(2,4,8: 51)$ & $24\langle1,1,2\rangle + 3\langle1,1,1\rangle$ & $(2,2,1: 4)$ & $4 \times 8 \times 8$ & 180 & 204 \\
\addlinespace[2ex]
\multirow{2}{*}{$(3,4,8: 75)$} & \multirow{2}{*}{$15\langle1,1,2\rangle + 15\langle1,1,3\rangle$} & $(4,4,1: 16)$ & $8 \times 12 \times 16$ & 960 & 1200 \\ 
 & & $(3,3,2: 15)$ & $9 \times 12 \times 16$ & 1035 & 1125 \\
\addlinespace[2ex]
\multirow{3}{*}{$(3,4,13: 123)$} & \multirow{3}{*}{$13\langle1, 1, 5\rangle + 7\langle1,1,4\rangle + 10\langle1,1,3\rangle$} & $(2,4,1: 8)$ & $6 \times 13 \times 16$ & 798 & 984 \\ 
 & & $(4,4,1: 16)$ & $12 \times 13 \times 16$ & 1509 & 1968 \\
 & & $(5,4,1: 20)$ & $13 \times 15 \times 16$ & 1885 & 2460 \\
\addlinespace[2ex]
$(3,5,5: 58)$ & $6\langle1,2,1\rangle + 46\langle1,1,1\rangle$ & $(3,2,3: 15)$ & $9 \times 10 \times 15$ & 864 & 870 \\
\addlinespace[0.5ex]\hline
\end{tabular}
\end{table}

\subsection{Searching for Alternative Schemes}

The framework also includes a tool for finding alternative schemes, i.e., schemes that are not equivalent in terms of their coefficients. For each scheme, a hash is computed based on the sorted list of coefficient strings (concatenated $u^{(i)}$, $v^{(i)}$, $w^{(i)}$). Only schemes with distinct hashes are retained. This prevents storing thousands of essentially identical schemes that differ only by a permutation of the rank components.

A more advanced mode searches for schemes with alternative structures. Instead of using coefficient strings, the hash is based on a randomly selected representation of the bud structure (the elementary tensor decomposition described in Section~\ref{subsec:tensor_structure}). This allows grouping schemes not by exact coefficients but by the pattern of buds they exhibit. Schemes that are coefficient-different but structurally similar can still be distinguished when needed.

This tool played a key role in discovering many of the improvements reported in this paper. In particular, for formats that already yield good results in serendipitous products, searching for alternative structures often revealed new schemes with richer bud configurations, enabling even larger rank savings.

%% file: structure/results.tex
\section{Results}
\label{sec:results}

\subsection{Experimental Overview}

Experiments were run on two machines: a Lenovo Legion laptop with a 20-core Intel Core Ultra 7 255HX and a desktop computer with a 12-core Intel Core i7-8700. The meta flip graph framework was executed with three different initialization strategies: (1) starting from a single naive scheme $(2,2,2:8)$; (2) starting from a set of naive schemes including $(2,2,3:12)$, $(2,2,4:16)$, $(2,3,3:18)$, $(2,3,4:24)$, $(2,3,5:30)$, $(2,4,4:32)$, $(3,3,3:27)$, $(3,3,4:36)$, $(3,3,5:45)$, $(3,4,4:48)$; and (3) starting from previously discovered in \cite{moosbauer2025flip} schemes $(5,5,5:93)$ and $(6,6,6:153)$ to replicate the original meta flip graph experiments~\cite{kauers2025exploring}. In typical runs, $16384$ parallel workers were used, with reset after $5$–$20$ million flips without improvement. Individual runs took from several days to two weeks. In total, approximately $25$ million distinct schemes were collected, of which about $4$ million have optimal ranks for their respective formats.

All collected schemes were analyzed for their bud structure and decomposed into elementary tensors as described in Section~\ref{subsec:tensor_structure}. The serendipitous product construction was then applied to all possible second schemes. To obtain ternary coefficients in the resulting constructions, all available schemes of optimal rank for the required auxiliary formats were tested, and those yielding ternary coefficients were preferred.

\subsection{Improved Ranks}

Compared to the previous state documented in~\cite{perminov2026fast}, ranks were improved for \improvedCount\ matrix multiplication formats. The vast majority of these improvements were obtained through the serendipitous product construction. Only about one tenth of them came directly from meta flip graph search alone. All improvements are listed in Table~\ref{tab:improved}. The column $\mathbb{F}$ indicates the coefficient ring used by the scheme (see Section~\ref{sec:preliminaries} for definitions), and the column $\omega$ is computed as $3 \log_{n m p}(r)$.

\begin{table}[ht!]
	\caption{Improved ranks for matrix multiplication formats}
    \label{tab:improved}
	\centering
    \scriptsize
	\begin{tabular}{ccccc|ccccc|ccccc}
        \toprule
        $n \times m \times p$ & $r_\text{prev}$ & $r_\text{new}$ & $\mathbb{F}$ & $\omega$ & $n \times m \times p$ & $r_\text{prev}$ & $r_\text{new}$ & $\mathbb{F}$ & $\omega$ & $n \times m \times p$ & $r_\text{prev}$ & $r_\text{new}$ & $\mathbb{F}$ & $\omega$ \\
        \midrule
$2 \times 4 \times 11$ & 71 & 70 & $\mathbb{Z}_T$ & 2.847 & $6 \times 8 \times 16$ & 511 & 510 & $\mathbb{Z}_T$ & 2.815 & $9 \times 9 \times 9$ & 498 & 486 & $\mathbb{Z}_T$ & 2.815 \\
$3 \times 5 \times 9$ & 104 & 102 & $\mathbb{Z}_T$ & 2.829 & $6 \times 9 \times 9$ & 342 & 332 & $\mathbb{Q}$ & 2.815 & $9 \times 9 \times 14$ & 726 & 720 & $\mathbb{Q}$ & 2.806 \\
$3 \times 5 \times 10$ & 115 & 114 & $\mathbb{Z}_T$ & 2.836 & $6 \times 9 \times 10$ & 373 & 367 & $\mathbb{Q}$ & 2.816 & $9 \times 9 \times 15$ & 783 & 760 & $\mathbb{Q}$ & 2.802 \\
$3 \times 7 \times 9$ & 142 & 141 & $\mathbb{Z}_T$ & 2.832 & $6 \times 9 \times 11$ & 407 & 404 & $\mathbb{Q}$ & 2.819 & $9 \times 9 \times 16$ & 825 & 822 & $\mathbb{Q}$ & 2.809 \\
$3 \times 7 \times 13$ & 205 & 204 & $\mathbb{Z}_T$ & 2.844 & $6 \times 9 \times 12$ & 434 & 429 & $\mathbb{Q}$ & 2.809 & $9 \times 10 \times 12$ & 684 & 668 & $\mathbb{Q}$ & 2.794 \\
$3 \times 7 \times 14$ & 220 & 219 & $\mathbb{Z}_T$ & 2.845 & $6 \times 9 \times 13$ & 474 & 468 & $\mathbb{Q}$ & 2.814 & $9 \times 10 \times 13$ & 765 & 758 & $\mathbb{Q}$ & 2.816 \\
$3 \times 7 \times 15$ & 236 & 235 & $\mathbb{Q}$ & 2.847 & $6 \times 9 \times 14$ & 500 & 494 & $\mathbb{Q}$ & 2.807 & $9 \times 10 \times 14$ & 819 & 808 & $\mathbb{Q}$ & 2.813 \\
$3 \times 9 \times 11$ & 224 & 222 & $\mathbb{Q}$ & 2.847 & $6 \times 9 \times 15$ & 532 & 529 & $\mathbb{Q}$ & 2.809 & $9 \times 10 \times 15$ & 870 & 864 & $\mathbb{Z}_T$ & 2.814 \\
$3 \times 9 \times 13$ & 262 & 261 & $\mathbb{Q}$ & 2.848 & $6 \times 9 \times 16$ & 556 & 552 & $\mathbb{Q}$ & 2.801 & $9 \times 10 \times 16$ & 930 & 916 & $\mathbb{Q}$ & 2.813 \\
$3 \times 9 \times 14$ & 283 & 281 & $\mathbb{Z}_T$ & 2.850 & $6 \times 10 \times 15$ & 597 & 594 & $\mathbb{Z}_T$ & 2.817 & $9 \times 11 \times 12$ & 754 & 738 & $\mathbb{Q}$ & 2.798 \\
$3 \times 10 \times 11$ & 249 & 248 & $\mathbb{Q}$ & 2.852 & $6 \times 11 \times 12$ & 524 & 521 & $\mathbb{Z}_T$ & 2.812 & $9 \times 11 \times 14$ & 889 & 882 & $\mathbb{Q}$ & 2.813 \\
$3 \times 10 \times 14$ & 314 & 312 & $\mathbb{Z}_T$ & 2.852 & $6 \times 11 \times 15$ & 661 & 653 & $\mathbb{Z}_T$ & 2.819 & $9 \times 11 \times 15$ & 960 & 956 & $\mathbb{Z}_T$ & 2.819 \\
$3 \times 10 \times 15$ & 336 & 335 & $\mathbb{Z}_T$ & 2.855 & $6 \times 11 \times 16$ & 695 & 684 & $\mathbb{Z}_T$ & 2.813 & $9 \times 11 \times 16$ & 1023 & 996 & $\mathbb{Z}_T$ & 2.811 \\
$3 \times 10 \times 16$ & 360 & 355 & $\mathbb{Z}_T$ & 2.853 & $6 \times 12 \times 13$ & 616 & 615 & $\mathbb{Z}_T$ & 2.816 & $9 \times 12 \times 13$ & 884 & 878 & $\mathbb{Z}_T$ & 2.806 \\
$3 \times 11 \times 14$ & 346 & 345 & $\mathbb{Q}$ & 2.857 & $6 \times 12 \times 14$ & 658 & 645 & $\mathbb{Q}$ & 2.806 & $9 \times 12 \times 14$ & 945 & 940 & $\mathbb{Q}$ & 2.805 \\
$3 \times 13 \times 13$ & 379 & 378 & $\mathbb{Q}$ & 2.859 & $6 \times 12 \times 15$ & 705 & 686 & $\mathbb{Q}$ & 2.805 & $9 \times 12 \times 15$ & 1000 & 996 & $\mathbb{Q}$ & 2.803 \\
$3 \times 13 \times 14$ & 408 & 407 & $\mathbb{Q}$ & 2.860 & $6 \times 12 \times 16$ & 746 & 736 & $\mathbb{Z}_T$ & 2.809 & $9 \times 12 \times 16$ & 1072 & 1035 & $\mathbb{Q}$ & 2.794 \\
$3 \times 13 \times 15$ & 436 & 435 & $\mathbb{Q}$ & 2.861 & $6 \times 13 \times 15$ & 771 & 763 & $\mathbb{Z}_T$ & 2.818 & $9 \times 13 \times 14$ & 1041 & 1024 & $\mathbb{Q}$ & 2.810 \\
$3 \times 13 \times 16$ & 465 & 464 & $\mathbb{Q}$ & 2.862 & $6 \times 13 \times 16$ & 819 & 798 & $\mathbb{Q}$ & 2.812 & $9 \times 13 \times 16$ & 1183 & 1167 & $\mathbb{Z}_T$ & 2.812 \\
$3 \times 14 \times 14$ & 440 & 438 & $\mathbb{Z}_T$ & 2.861 & $6 \times 14 \times 14$ & 777 & 776 & $\mathbb{Q}$ & 2.824 & $9 \times 14 \times 14$ & 1121 & 1101 & $\mathbb{Q}$ & 2.811 \\
$3 \times 14 \times 15$ & 470 & 469 & $\mathbb{Q}$ & 2.863 & $6 \times 14 \times 15$ & 825 & 814 & $\mathbb{Z}_T$ & 2.816 & $9 \times 14 \times 15$ & 1185 & 1175 & $\mathbb{Q}$ & 2.811 \\
$3 \times 14 \times 16$ & 502 & 500 & $\mathbb{Z}_T$ & 2.864 & $6 \times 14 \times 16$ & 880 & 864 & $\mathbb{Z}_T$ & 2.816 & $9 \times 14 \times 16$ & 1260 & 1254 & $\mathbb{Q}$ & 2.813 \\
$3 \times 15 \times 16$ & 536 & 534 & $\mathbb{Z}_T$ & 2.864 & $6 \times 15 \times 15$ & 870 & 859 & $\mathbb{Q}$ & 2.812 & $9 \times 15 \times 15$ & 1284 & 1236 & $\mathbb{Q}$ & 2.805 \\
$3 \times 16 \times 16$ & 574 & 569 & $\mathbb{Q}$ & 2.865 & $6 \times 15 \times 16$ & 928 & 920 & $\mathbb{Z}_T$ & 2.815 & $9 \times 15 \times 16$ & 1341 & 1320 & $\mathbb{Z}_T$ & 2.808 \\
$4 \times 4 \times 11$ & 130 & 129 & $\mathbb{Z}_T$ & 2.820 & $6 \times 16 \times 16$ & 988 & 972 & $\mathbb{Z}_T$ & 2.813 & $9 \times 16 \times 16$ & 1431 & 1380 & $\mathbb{Z}_T$ & 2.801 \\
$4 \times 6 \times 16$ & 280 & 276 & $\mathbb{Z}_T$ & 2.834 & $7 \times 8 \times 9$ & 350 & 347 & $\mathbb{Z}_T$ & 2.820 & $10 \times 10 \times 12$ & 770 & 766 & $\mathbb{Z}_T$ & 2.810 \\
$4 \times 7 \times 8$ & 164 & 161 & $\mathbb{Z}_T$ & 2.817 & $7 \times 8 \times 12$ & 454 & 452 & $\mathbb{Z}_T$ & 2.817 & $10 \times 11 \times 12$ & 850 & 849 & $\mathbb{Q}$ & 2.816 \\
$4 \times 7 \times 11$ & 226 & 224 & $\mathbb{Q}$ & 2.833 & $7 \times 8 \times 15$ & 570 & 557 & $\mathbb{Q}$ & 2.817 & $10 \times 11 \times 15$ & 1055 & 1050 & $\mathbb{Z}_T$ & 2.817 \\
$4 \times 7 \times 12$ & 246 & 242 & $\mathbb{Z}_T$ & 2.831 & $7 \times 8 \times 16$ & 603 & 598 & $\mathbb{Z}_T$ & 2.822 & $10 \times 12 \times 12$ & 910 & 902 & $\mathbb{Z}_T$ & 2.807 \\
$4 \times 7 \times 13$ & 266 & 265 & $\mathbb{Z}_T$ & 2.839 & $7 \times 9 \times 9$ & 398 & 396 & $\mathbb{Z}_T$ & 2.830 & $10 \times 12 \times 15$ & 1130 & 1122 & $\mathbb{Z}_T$ & 2.811 \\
$4 \times 7 \times 14$ & 285 & 284 & $\mathbb{Z}_T$ & 2.838 & $7 \times 9 \times 10$ & 437 & 433 & $\mathbb{Z}_T$ & 2.825 & $10 \times 12 \times 16$ & 1190 & 1176 & $\mathbb{Q}$ & 2.805 \\
$4 \times 7 \times 15$ & 307 & 305 & $\mathbb{Z}_T$ & 2.841 & $7 \times 9 \times 12$ & 510 & 508 & $\mathbb{Q}$ & 2.820 & $10 \times 13 \times 15$ & 1242 & 1230 & $\mathbb{Z}_T$ & 2.818 \\
$4 \times 7 \times 16$ & 324 & 322 & $\mathbb{Z}_T$ & 2.838 & $7 \times 10 \times 12$ & 564 & 557 & $\mathbb{Q}$ & 2.817 & $10 \times 13 \times 16$ & 1326 & 1318 & $\mathbb{Q}$ & 2.821 \\
$4 \times 8 \times 8$ & 182 & 180 & $\mathbb{Z}_T$ & 2.809 & $7 \times 10 \times 15$ & 703 & 694 & $\mathbb{Z}_T$ & 2.821 & $10 \times 14 \times 15$ & 1327 & 1314 & $\mathbb{Z}_T$ & 2.816 \\
$4 \times 9 \times 10$ & 255 & 250 & $\mathbb{Z}_T$ & 2.814 & $7 \times 10 \times 16$ & 742 & 736 & $\mathbb{Q}$ & 2.821 & $10 \times 14 \times 16$ & 1423 & 1398 & $\mathbb{Q}$ & 2.817 \\
$4 \times 9 \times 11$ & 279 & 275 & $\mathbb{Z}_T$ & 2.817 & $7 \times 12 \times 15$ & 831 & 815 & $\mathbb{Z}_T$ & 2.817 & $10 \times 15 \times 15$ & 1395 & 1385 & $\mathbb{Q}$ & 2.811 \\
$4 \times 9 \times 13$ & 329 & 325 & $\mathbb{Z}_T$ & 2.822 & $7 \times 12 \times 16$ & 880 & 878 & $\mathbb{Z}_T$ & 2.823 & $10 \times 15 \times 16$ & 1497 & 1482 & $\mathbb{Q}$ & 2.814 \\
$4 \times 9 \times 14$ & 355 & 350 & $\mathbb{Z}_T$ & 2.824 & $7 \times 13 \times 16$ & 968 & 962 & $\mathbb{Q}$ & 2.829 & $10 \times 16 \times 16$ & 1586 & 1560 & $\mathbb{Q}$ & 2.811 \\
$4 \times 9 \times 16$ & 400 & 398 & $\mathbb{Z}_T$ & 2.826 & $7 \times 14 \times 15$ & 969 & 952 & $\mathbb{Z}_T$ & 2.821 & $11 \times 11 \times 12$ & 936 & 922 & $\mathbb{Q}$ & 2.813 \\
$4 \times 10 \times 15$ & 417 & 413 & $\mathbb{Q}$ & 2.825 & $7 \times 14 \times 16$ & 1034 & 1022 & $\mathbb{Q}$ & 2.825 & $11 \times 12 \times 12$ & 990 & 968 & $\mathbb{Q}$ & 2.799 \\
$4 \times 11 \times 12$ & 365 & 362 & $\mathbb{Z}_T$ & 2.819 & $7 \times 15 \times 16$ & 1099 & 1083 & $\mathbb{Q}$ & 2.823 & $11 \times 12 \times 13$ & 1092 & 1082 & $\mathbb{Q}$ & 2.814 \\
$4 \times 11 \times 15$ & 452 & 449 & $\mathbb{Q}$ & 2.822 & $8 \times 8 \times 15$ & 635 & 628 & $\mathbb{Q}$ & 2.815 & $11 \times 12 \times 14$ & 1182 & 1153 & $\mathbb{Q}$ & 2.812 \\
$4 \times 11 \times 16$ & 489 & 480 & $\mathbb{Z}_T$ & 2.825 & $8 \times 8 \times 16$ & 671 & 666 & $\mathbb{Z}_T$ & 2.814 & $11 \times 12 \times 15$ & 1240 & 1234 & $\mathbb{Z}_T$ & 2.813 \\
$4 \times 12 \times 12$ & 390 & 389 & $\mathbb{Z}_T$ & 2.815 & $8 \times 9 \times 10$ & 487 & 482 & $\mathbb{Z}_T$ & 2.817 & $11 \times 12 \times 16$ & 1312 & 1278 & $\mathbb{Q}$ & 2.803 \\
$4 \times 12 \times 13$ & 426 & 422 & $\mathbb{Q}$ & 2.818 & $8 \times 9 \times 11$ & 531 & 521 & $\mathbb{Z}_T$ & 2.812 & $11 \times 13 \times 15$ & 1377 & 1371 & $\mathbb{Q}$ & 2.825 \\
$4 \times 12 \times 14$ & 456 & 452 & $\mathbb{Q}$ & 2.817 & $8 \times 9 \times 13$ & 624 & 615 & $\mathbb{Z}_T$ & 2.816 & $11 \times 13 \times 16$ & 1452 & 1446 & $\mathbb{Q}$ & 2.822 \\
$4 \times 12 \times 16$ & 520 & 513 & $\mathbb{Z}_T$ & 2.818 & $8 \times 9 \times 14$ & 666 & 654 & $\mathbb{Z}_T$ & 2.812 & $11 \times 14 \times 15$ & 1460 & 1432 & $\mathbb{Z}_T$ & 2.815 \\
$4 \times 13 \times 15$ & 528 & 520 & $\mathbb{Q}$ & 2.817 & $8 \times 9 \times 15$ & 705 & 699 & $\mathbb{Z}_T$ & 2.813 & $11 \times 14 \times 16$ & 1548 & 1520 & $\mathbb{Q}$ & 2.814 \\
$4 \times 13 \times 16$ & 568 & 560 & $\mathbb{Z}_T$ & 2.823 & $8 \times 9 \times 16$ & 746 & 735 & $\mathbb{Z}_T$ & 2.809 & $11 \times 15 \times 16$ & 1656 & 1605 & $\mathbb{Q}$ & 2.811 \\
$4 \times 14 \times 15$ & 568 & 557 & $\mathbb{Q}$ & 2.817 & $8 \times 10 \times 10$ & 532 & 528 & $\mathbb{Z}_T$ & 2.814 & $12 \times 12 \times 13$ & 1152 & 1144 & $\mathbb{Q}$ & 2.804 \\
$4 \times 14 \times 16$ & 610 & 598 & $\mathbb{Z}_T$ & 2.822 & $8 \times 10 \times 12$ & 630 & 624 & $\mathbb{Q}$ & 2.812 & $12 \times 12 \times 14$ & 1240 & 1234 & $\mathbb{Q}$ & 2.806 \\
$4 \times 15 \times 15$ & 600 & 596 & $\mathbb{Q}$ & 2.818 & $8 \times 10 \times 14$ & 728 & 726 & $\mathbb{Z}_T$ & 2.815 & $12 \times 12 \times 16$ & 1392 & 1380 & $\mathbb{Z}_T$ & 2.801 \\
$4 \times 15 \times 16$ & 640 & 632 & $\mathbb{Q}$ & 2.817 & $8 \times 10 \times 15$ & 784 & 778 & $\mathbb{Q}$ & 2.817 & $12 \times 13 \times 14$ & 1382 & 1370 & $\mathbb{Q}$ & 2.818 \\
$4 \times 16 \times 16$ & 676 & 666 & $\mathbb{Z}_T$ & 2.814 & $8 \times 10 \times 16$ & 826 & 822 & $\mathbb{Q}$ & 2.814 & $12 \times 13 \times 15$ & 1460 & 1442 & $\mathbb{Q}$ & 2.813 \\
$5 \times 5 \times 12$ & 208 & 204 & $\mathbb{Z}_T$ & 2.797 & $8 \times 11 \times 12$ & 680 & 676 & $\mathbb{Q}$ & 2.808 & $12 \times 13 \times 16$ & 1548 & 1509 & $\mathbb{Q}$ & 2.807 \\
$5 \times 5 \times 13$ & 228 & 227 & $\mathbb{Z}_T$ & 2.814 & $8 \times 11 \times 15$ & 859 & 848 & $\mathbb{Q}$ & 2.815 & $12 \times 14 \times 14$ & 1481 & 1449 & $\mathbb{Q}$ & 2.813 \\
$5 \times 5 \times 14$ & 248 & 244 & $\mathbb{Z}_T$ & 2.815 & $8 \times 11 \times 16$ & 914 & 904 & $\mathbb{Q}$ & 2.817 & $12 \times 14 \times 15$ & 1540 & 1538 & $\mathbb{Q}$ & 2.811 \\
$5 \times 5 \times 15$ & 266 & 262 & $\mathbb{Z}_T$ & 2.818 & $8 \times 12 \times 13$ & 798 & 781 & $\mathbb{Q}$ & 2.803 & $12 \times 14 \times 16$ & 1638 & 1617 & $\mathbb{Q}$ & 2.807 \\
$5 \times 5 \times 16$ & 284 & 280 & $\mathbb{Z}_T$ & 2.821 & $8 \times 12 \times 14$ & 861 & 843 & $\mathbb{Q}$ & 2.806 & $12 \times 15 \times 16$ & 1728 & 1725 & $\mathbb{Z}_T$ & 2.807 \\
$5 \times 6 \times 10$ & 217 & 216 & $\mathbb{Z}_T$ & 2.827 & $8 \times 12 \times 15$ & 915 & 904 & $\mathbb{Q}$ & 2.808 & $12 \times 16 \times 16$ & 1824 & 1815 & $\mathbb{Q}$ & 2.803 \\
$5 \times 9 \times 15$ & 474 & 463 & $\mathbb{Q}$ & 2.826 & $8 \times 13 \times 14$ & 945 & 944 & $\mathbb{Q}$ & 2.822 & $13 \times 14 \times 15$ & 1698 & 1681 & $\mathbb{Z}_T$ & 2.816 \\
$5 \times 10 \times 12$ & 413 & 408 & $\mathbb{Z}_T$ & 2.819 & $8 \times 13 \times 15$ & 1005 & 991 & $\mathbb{Z}_T$ & 2.815 & $13 \times 14 \times 16$ & 1806 & 1800 & $\mathbb{Q}$ & 2.819 \\
$5 \times 11 \times 12$ & 455 & 454 & $\mathbb{Q}$ & 2.827 & $8 \times 13 \times 16$ & 1064 & 1054 & $\mathbb{Q}$ & 2.815 & $13 \times 15 \times 15$ & 1803 & 1797 & $\mathbb{Z}_T$ & 2.817 \\
$5 \times 12 \times 15$ & 615 & 612 & $\mathbb{Z}_T$ & 2.830 & $8 \times 14 \times 14$ & 1008 & 1004 & $\mathbb{Q}$ & 2.818 & $13 \times 15 \times 16$ & 1908 & 1885 & $\mathbb{Z}_T$ & 2.812 \\
$5 \times 12 \times 16$ & 656 & 655 & $\mathbb{Q}$ & 2.833 & $8 \times 14 \times 15$ & 1080 & 1063 & $\mathbb{Z}_T$ & 2.815 & $14 \times 14 \times 15$ & 1813 & 1798 & $\mathbb{Z}_T$ & 2.815 \\
$5 \times 15 \times 15$ & 762 & 761 & $\mathbb{Z}_T$ & 2.833 & $8 \times 14 \times 16$ & 1138 & 1104 & $\mathbb{Q}$ & 2.806 & $14 \times 14 \times 16$ & 1938 & 1931 & $\mathbb{Q}$ & 2.819 \\
$6 \times 6 \times 13$ & 316 & 315 & $\mathbb{Q}$ & 2.807 & $8 \times 15 \times 15$ & 1140 & 1130 & $\mathbb{Q}$ & 2.814 & $14 \times 15 \times 15$ & 1905 & 1890 & $\mathbb{Q}$ & 2.810 \\
$6 \times 7 \times 9$ & 268 & 264 & $\mathbb{Z}_T$ & 2.819 & $8 \times 15 \times 16$ & 1198 & 1185 & $\mathbb{Q}$ & 2.809 & $14 \times 16 \times 16$ & 2142 & 2128 & $\mathbb{Q}$ & 2.809 \\
$6 \times 8 \times 10$ & 329 & 327 & $\mathbb{Z}_T$ & 2.813 & $8 \times 16 \times 16$ & 1248 & 1230 & $\mathbb{Q}$ & 2.799 & $15 \times 15 \times 16$ & 2155 & 2132 & $\mathbb{Z}_T$ & 2.808 \\
\hline
\end{tabular}
\end{table}

\subsection{Rediscovered Ternary Schemes}

For \rediscoveredZT\ formats, a scheme with $\mathbb{Z}_T$ coefficients was found where previously only schemes over $\mathbb{Z}$ or $\mathbb{Q}$ were known. The majority of these rediscoveries were obtained directly through meta flip graph search in $\mathbb{Z}_T$ or $\mathbb{Z}_2$, followed by Hensel lifting and rational reconstruction. Table~\ref{tab:rediscovered} lists these formats together with the previously known coefficient ring ($\mathbb{F}$).

\begin{table}[ht!]
	\caption{Rediscovered schemes with ternary coefficients ($\mathbb{Z}_T$)}
    \label{tab:rediscovered}
	\centering
    \small
	\begin{tabular}{ccc|ccc|ccc|ccc}
        \toprule
		$n \times m \times p$ & $r$ & $\mathbb{F}_\text{known}$ & $n \times m \times p$ & $r$ & $\mathbb{F}_\text{known}$ & $n \times m \times p$ & $r$ & $\mathbb{F}_\text{known}$ & $n \times m \times p$ & $r$ & $\mathbb{F}_\text{known}$ \\
        \midrule
$2 \times 4 \times 9$ & 58 & $\mathbb{Q}$ & $2 \times 7 \times 11$ & 121 & $\mathbb{Z}$ & $3 \times 5 \times 13$ & 147 & $\mathbb{Z}$ & $5 \times 9 \times 12$ & 377 & $\mathbb{Z}$ \\
$2 \times 4 \times 10$ & 64 & $\mathbb{Q}$ & $2 \times 7 \times 12$ & 131 & $\mathbb{Z}$ & $3 \times 5 \times 14$ & 158 & $\mathbb{Z}$ & $5 \times 10 \times 11$ & 386 & $\mathbb{Z}$ \\
$2 \times 4 \times 13$ & 83 & $\mathbb{Q}$ & $2 \times 7 \times 13$ & 142 & $\mathbb{Z}$ & $3 \times 5 \times 15$ & 169 & $\mathbb{Z}$ & $5 \times 10 \times 13$ & 451 & $\mathbb{Z}$ \\
$2 \times 5 \times 7$ & 55 & $\mathbb{Z}$ & $2 \times 7 \times 14$ & 152 & $\mathbb{Z}$ & $3 \times 5 \times 16$ & 180 & $\mathbb{Z}$ & $5 \times 10 \times 14$ & 481 & $\mathbb{Z}$ \\
$2 \times 5 \times 8$ & 63 & $\mathbb{Z}$ & $2 \times 7 \times 15$ & 164 & $\mathbb{Z}$ & $3 \times 7 \times 7$ & 111 & $\mathbb{Z}$ & $5 \times 10 \times 15$ & 519 & $\mathbb{Z}$ \\
$2 \times 5 \times 10$ & 79 & $\mathbb{Q}$ & $2 \times 7 \times 16$ & 175 & $\mathbb{Z}$ & $3 \times 8 \times 9$ & 163 & $\mathbb{Z}$ & $5 \times 10 \times 16$ & 549 & $\mathbb{Z}$ \\
$2 \times 5 \times 13$ & 102 & $\mathbb{Z}$ & $2 \times 8 \times 9$ & 113 & $\mathbb{Z}$ & $3 \times 8 \times 10$ & 180 & $\mathbb{Z}$ & $5 \times 11 \times 16$ & 609 & $\mathbb{Z}$ \\
$2 \times 5 \times 14$ & 110 & $\mathbb{Z}$ & $2 \times 8 \times 10$ & 125 & $\mathbb{Z}$ & $3 \times 8 \times 11$ & 198 & $\mathbb{Z}$ & $5 \times 14 \times 14$ & 672 & $\mathbb{Z}$ \\
$2 \times 5 \times 15$ & 118 & $\mathbb{Z}$ & $2 \times 8 \times 11$ & 138 & $\mathbb{Z}$ & $3 \times 8 \times 15$ & 270 & $\mathbb{Z}$ & $5 \times 15 \times 16$ & 813 & $\mathbb{Z}$ \\
$2 \times 5 \times 16$ & 126 & $\mathbb{Z}$ & $2 \times 8 \times 12$ & 150 & $\mathbb{Z}$ & $3 \times 8 \times 16$ & 288 & $\mathbb{Z}$ & $6 \times 10 \times 11$ & 446 & $\mathbb{Z}$ \\
$2 \times 6 \times 6$ & 56 & $\mathbb{Z}$ & $2 \times 8 \times 13$ & 163 & $\mathbb{Z}$ & $3 \times 11 \times 11$ & 274 & $\mathbb{Q}$ & $6 \times 10 \times 12$ & 476 & $\mathbb{Z}$ \\
$2 \times 6 \times 7$ & 66 & $\mathbb{Z}$ & $2 \times 8 \times 14$ & 175 & $\mathbb{Z}$ & $4 \times 10 \times 13$ & 361 & $\mathbb{Z}$ & $6 \times 10 \times 13$ & 520 & $\mathbb{Z}$ \\
$2 \times 6 \times 8$ & 75 & $\mathbb{Z}$ & $2 \times 10 \times 15$ & 234 & $\mathbb{Z}$ & $4 \times 10 \times 14$ & 385 & $\mathbb{Z}$ & $6 \times 10 \times 14$ & 553 & $\mathbb{Z}$ \\
$2 \times 6 \times 11$ & 103 & $\mathbb{Z}$ & $3 \times 3 \times 10$ & 69 & $\mathbb{Q}$ & $4 \times 10 \times 16$ & 441 & $\mathbb{Z}$ & $6 \times 10 \times 16$ & 630 & $\mathbb{Z}$ \\
$2 \times 6 \times 12$ & 112 & $\mathbb{Z}$ & $3 \times 3 \times 11$ & 76 & $\mathbb{Q}$ & $4 \times 11 \times 11$ & 340 & $\mathbb{Z}$ & $7 \times 10 \times 11$ & 526 & $\mathbb{Z}$ \\
$2 \times 6 \times 13$ & 122 & $\mathbb{Z}$ & $3 \times 4 \times 8$ & 73 & $\mathbb{Z}$ & $4 \times 11 \times 14$ & 429 & $\mathbb{Z}$ & $7 \times 10 \times 13$ & 614 & $\mathbb{Z}$ \\
$2 \times 6 \times 14$ & 131 & $\mathbb{Z}$ & $3 \times 5 \times 6$ & 68 & $\mathbb{Z}$ & $4 \times 14 \times 14$ & 532 & $\mathbb{Z}$ & $7 \times 10 \times 14$ & 653 & $\mathbb{Z}$ \\
$2 \times 6 \times 16$ & 150 & $\mathbb{Z}$ & $3 \times 5 \times 7$ & 79 & $\mathbb{Z}$ & $5 \times 7 \times 9$ & 229 & $\mathbb{Z}$ & $7 \times 11 \times 16$ & 822 & $\mathbb{Z}$ \\
$2 \times 7 \times 7$ & 76 & $\mathbb{Z}$ & $3 \times 5 \times 8$ & 90 & $\mathbb{Z}$ & $5 \times 8 \times 9$ & 260 & $\mathbb{Z}$ & $7 \times 14 \times 14$ & 909 & $\mathbb{Z}$ \\
$2 \times 7 \times 9$ & 99 & $\mathbb{Z}$ & $3 \times 5 \times 11$ & 126 & $\mathbb{Z}$ & $5 \times 8 \times 16$ & 445 & $\mathbb{Z}$ & $8 \times 12 \times 16$ & 960 & $\mathbb{Q}$ \\
$2 \times 7 \times 10$ & 110 & $\mathbb{Z}$ & $3 \times 5 \times 12$ & 136 & $\mathbb{Z}$ & $5 \times 9 \times 11$ & 353 & $\mathbb{Z}$ & $9 \times 13 \times 15$ & 1119 & $\mathbb{Z}$ \\
        \bottomrule
	\end{tabular}
\end{table}

\subsection{Schemes Beating Strassen's Exponent}

A total of \betterStrassen\ new schemes with asymptotic exponent $\omega < \log_2 7 \approx 2.80735$ were discovered among the 680 formats considered, bringing the overall count of such schemes (within this set) to \betterStrassenTotal. Among these, five have $\mathbb{Z}_T$ coefficients. All such new schemes are listed in Table~\ref{tab:better}.

\begin{table}[ht!]
	\caption{New discovered schemes with asymptotic exponent better than Strassen's}
    \label{tab:better}
	\centering
    \small
	\begin{tabular}{ccccc|ccccc}
        \toprule
		$n \times m \times p$ & $r$ & $\mathbb{F}$ & $\omega_{\text{prev}}$ & $\omega_{\text{new}}$ &  $n \times m \times p$ & $r$ & $\mathbb{F}$ & $\omega_{\text{prev}}$ & $\omega_{\text{new}}$ \\
        \midrule
$5 \times 5 \times 12$ & 204 & $\mathbb{Z}_T$ & 2.80737 & 2.79715 & $9 \times 12 \times 16$ & 1035 & $\mathbb{Q}$ & 2.80786 & 2.79373 \\
$6 \times 6 \times 13$ & 315 & $\mathbb{Q}$ & 2.80838 & 2.80683 & $9 \times 15 \times 15$ & 1236 & $\mathbb{Q}$ & 2.82048 & 2.80546 \\
$6 \times 12 \times 14$ & 645 & $\mathbb{Q}$ & 2.81498 & 2.80632 & $9 \times 16 \times 16$ & 1380 & $\mathbb{Z}_T$ & 2.81546 & 2.80139 \\
$6 \times 12 \times 15$ & 686 & $\mathbb{Q}$ & 2.81681 & 2.80507 & $10 \times 12 \times 12$ & 902 & $\mathbb{Z}_T$ & 2.81067 & 2.80703 \\
$8 \times 12 \times 13$ & 781 & $\mathbb{Q}$ & 2.81182 & 2.80276 & $10 \times 12 \times 16$ & 1176 & $\mathbb{Q}$ & 2.81017 & 2.80548 \\
$8 \times 12 \times 14$ & 843 & $\mathbb{Q}$ & 2.81454 & 2.80574 & $11 \times 12 \times 12$ & 968 & $\mathbb{Q}$ & 2.80862 & 2.79947 \\
$8 \times 14 \times 16$ & 1104 & $\mathbb{Q}$ & 2.81816 & 2.80601 & $11 \times 12 \times 16$ & 1278 & $\mathbb{Q}$ & 2.81343 & 2.80314 \\
$9 \times 9 \times 14$ & 720 & $\mathbb{Q}$ & 2.80979 & 2.80625 & $12 \times 12 \times 14$ & 1234 & $\mathbb{Q}$ & 2.80838 & 2.80647 \\
$9 \times 9 \times 15$ & 760 & $\mathbb{Q}$ & 2.81442 & 2.80182 & $12 \times 13 \times 16$ & 1509 & $\mathbb{Q}$ & 2.81679 & 2.80700 \\
$9 \times 11 \times 12$ & 738 & $\mathbb{Q}$ & 2.80736 & 2.79827 & $12 \times 14 \times 16$ & 1617 & $\mathbb{Q}$ & 2.81182 & 2.80692 \\
$9 \times 12 \times 13$ & 878 & $\mathbb{Z}_T$ & 2.80849 & 2.80567 & $12 \times 15 \times 16$ & 1725 & $\mathbb{Z}_T$ & 2.80761 & 2.80696 \\
$9 \times 12 \times 14$ & 940 & $\mathbb{Q}$ & 2.80741 & 2.80523 &  \\
        \bottomrule
	\end{tabular}
\end{table}

\subsection{Overall Statistics}

Across all 680 formats $n \times m \times p$ with $2 \le n \le m \le p \le 16$, the distribution of coefficient rings is as follows:

\begin{itemize}
    \item $\mathbb{Z}_T$: \countZT\ schemes (\percentZT\%);
    \item $\mathbb{Z}$: \countZ\ schemes (\percentZ\%);
    \item $\mathbb{Q}$: \countQ\ schemes (\percentQ\%).
\end{itemize}

The total number of formats with $\omega < \log_2 7$ is now \betterStrassenTotal, an increase of \betterStrassen\ from the previous count of 29 reported in~\cite{perminov2026fast}.

%% file: structure/discussion.tex
\section{Discussion and Future Work}
\label{sec:discussion}

\subsection{New Meta-Operators}

The current meta flip graph framework supports four dimension-changing operations: \texttt{project}, \texttt{extend}, \texttt{merge}, and \texttt{product}. Adding block matrix multiplication and serendipitous product as explicit meta-operators could further expand the reachable space of schemes. For larger formats (beyond $6 \times 6 \times 6$), naive \texttt{extend} and \texttt{merge} often produce ranks far from optimal, whereas a serendipitous product operator would directly construct schemes with better ranks by exploiting buds structure. Implementing such operators and integrating them into the random walk selection process is a promising direction.

\subsection{Searching for Bud-Rich Schemes}

Schemes with higher rank tend to contain more buds than optimal-rank schemes. The meta flip graph framework discovered many schemes with ranks several steps above the optimum, and these were all examined for serendipitous products. Unfortunately, optimal-rank schemes typically have very few buds. A promising direction for future work is to search specifically for alternative schemes of optimal (or near-optimal) rank that are rich in buds, rather than relying on higher-rank schemes.

\subsection{Instability of Hensel Lifting}

Hensel lifting is highly sensitive due to the rectangular nature of the Brent equations. A single flip in $\mathbb{Z}_2$ or $\mathbb{Z}_3$ can turn a ternary scheme into an irreducible rational scheme, whereas the same flip performed directly in $\mathbb{Z}_T$ would preserve ternarity. For small formats, an SMT solver can be used to recover ternary coefficients. For larger formats, it is better not to discard a rational scheme immediately. Instead, the scheme can be flipped for a limited number of steps. Experiments show that this often yields an integer or ternary scheme within a few hundred flips.

\subsection{Towards Integer and Ternary Coefficients}

Many schemes discovered by meta flip graph search (in $\mathbb{Z}_2$ or $\mathbb{Z}_3$) lift to rational coefficients rather than integers or ternary values. However, empirical observations suggest that applying suitable transformations (such as rescalings and flips) can often eliminate fractions, yielding an equivalent scheme over $\mathbb{Z}$ or even $\mathbb{Z}_T$. Recent work by Morán et al.~\cite{moran2026complex} investigated the translation of schemes from complex coefficients to rational ones and provided criteria for when integer schemes cannot exist. A systematic study of which rational schemes admit equivalent integer or ternary representations remains an open problem. The flip graph framework, combined with Hensel lifting and rational reconstruction, appears well-suited for this question, and the collection of schemes obtained in the present work provides a rich source for future investigations.

\subsection{Reducing Arithmetic Complexity}

While rank reduction directly lowers the number of multiplications, the additive complexity (number of additions and subtractions) also significantly affects the practical performance of matrix multiplication schemes. Minimizing additions is a natural next step after rank optimization.

The problem of minimizing additive complexity is challenging. Common subexpression elimination (CSE) alone is not sufficient for solving it effectively. Recent work by Mårtensson et al.~\cite{maartensson2026and} describes various techniques for reducing additive complexity and discusses how to integrate addition reduction into the scheme generation pipeline. Other tools have been developed specifically for this task. The PLinOpt framework~\cite{dumas2024plinopt, dumas2025towards} provides optimization of linear, bilinear, and trilinear programs, and has been applied to reduce additive complexity for various matrix multiplication formats. A parallel heuristic search method~\cite{perminov2025parallel} explores the space of CSE strategies through random substitutions and sharing of promising partial solutions, achieving lower addition counts on a wide range of ternary schemes.

The Sedoglavic catalog~\cite{sedoglavic2025yet} focuses exclusively on ranks and does not track additive complexity. A systematic study of additive complexity across many formats is still missing, largely because the problem is difficult, especially for larger schemes. A valuable direction for future work would be to compute, or at least estimate, the minimal additive complexity for formats up to some moderate bound, say $16 \times 16 \times 16$. The large collection of schemes obtained in the present work -- many of which have optimal or near-optimal ranks -- provides a rich starting point for such an investigation. Applying CSE and other optimization techniques to these schemes could reveal trade-offs between multiplicative and additive complexity and potentially identify schemes that are optimal in both respects.

\subsection{Numerical Stability and Practical Efficiency}

All schemes presented in this work are exact algebraic identities over the rational numbers (or integers). No analysis of numerical stability has been performed. For floating-point implementations, different schemes with the same rank may exhibit drastically different error accumulation behavior. Factors such as the magnitude of coefficients, the number of additions, and the specific cancellation patterns all affect stability. Similarly, practical efficiency depends not only on the rank but also on memory access patterns, parallelism, and vectorization opportunities. Evaluating and comparing schemes with respect to these criteria is an important direction for future work. The large collection of schemes obtained here provides a valuable dataset for such investigations, which could ultimately lead to practical algorithms that are both fast in theory and stable in practice.

%% file: structure/conclusion.tex
\section{Conclusion}
\label{sec:conclusion}

This paper demonstrated that combining the meta flip graph framework with the serendipitous product construction yields significant improvements in the search for fast matrix multiplication schemes. The open-source implementation of the ternary flip graph framework~\cite{perminov2026fast} was extended to support formats up to $16 \times 16 \times 16$, enabling exhaustive coverage of all 680 rectangular formats with $2 \le n \le m \le p \le 16$.

The main results are as follows:
\begin{itemize}
    \item Ranks were improved for \improvedCount\ formats compared to the previous state of the art.
    \item For \rediscoveredZT\ formats, ternary ($\mathbb{Z}_T$) schemes were found where previously only integer ($\mathbb{Z}$) or rational ($\mathbb{Q}$) schemes were known.
    \item A total of \betterStrassen\ new schemes with asymptotic exponent $\omega < \log_2 7$ were discovered, bringing the total number of such schemes to 51.
    \item The overall distribution of coefficient types across all 680 formats is now \countZT\ ternary (\percentZT\%), \countZ\ integer (\percentZ\%), and \countQ\ rational (\percentQ\%).
\end{itemize}

The serendipitous product proved to be a powerful tool for constructing larger formats with improved ranks, especially when combined with random walks that refine bud configurations. A particularly interesting observation is that the optimal scheme for a given format is not always the best candidate for serendipitous constructions: schemes with slightly higher ranks often possess richer bud structures that lead to greater overall savings.

The meta flip graph framework, with its ability to navigate across different dimensions and ranks, successfully generated a diverse collection of schemes that served as a rich source of buds. The combination of exhaustive search in small formats, random walks guided by rank-based selection, and systematic application of meta-operators proved effective for exploring the space of matrix multiplication algorithms.

All code and discovered schemes are available as open source at:
\begin{itemize}
    \item Flip graph framework: \url{https://github.com/dronperminov/ternary_flip_graph}
    \item Schemes and results: \url{https://github.com/dronperminov/FastMatrixMultiplication}
\end{itemize}